\documentclass[12pt]{article}
\usepackage{amssymb,amsmath,amsthm}
\usepackage{latexsym}

\newcommand{\beg}{\begin{equation}}
\newcommand{\ene}{\end{equation}}

\begin{document}
\title{
\textsc{\bf Quantum Computing and }\\
\textsc{\bf Shor`s  Factoring Algorithm} \footnote{Lectures at the
Volterra--CIRM International School "Quantum Computer and  Quantum
Information", Trento, Italy, July 25--31, 2001. }
\\
$~$\\}
\author{
\textsf{I.V. Volovich}
\\
\emph{Steklov Mathematical Institute}\\
\emph {Russian Academy of Sciences}\\
\emph{Gubkin St. 8, 117866, GSP-1, Moscow, Russia}\\
\emph{e-mail: volovich@mi.ras.ru}
}

\date {~}
\maketitle
\thispagestyle{empty}

\begin{abstract}
Lectures on  quantum computing. Contents:
Algorithms. Quantum circuits. Quantum Fourier transform.
Elements of number theory. Modular exponentiation.
Shor`s algorithm for finding the order.
Computational complexity of Schor`s algorithm.
Factoring integers. NP-complete problems.
\end{abstract}

\newpage
\tableofcontents

\section{Introduction}
\label{intro}

In these lectures a brief introduction to quantum computing and number
theory is given and
 Shor`s  algorithm  for factoring integers is described.
The lectures are based on the material from the forthcoming book
\cite{OV}.

Let us discuss the problem of factoring.
 It is known that every integer $N$ is uniquely
decomposable into a product
of prime numbers. However we do not know {\it efficient}
 (i.e. polynomial in
the number of operations) classical algorithms for factoring.
 Given a large integer $N,$ one
has to find efficiently  such integers $p$
and $q$ that $N=pq$ or to prove that such a factoring does not exist.
It is assumed that $p$ and $q$ are not equal to 1.

An algorithm of factoring the  number $N$ is called efficient
if the number of elementary arithmetical operations
which it uses for large $N$ is bounded by  a polynomial
in $n$ where  $n=\log N$ is the number of digits
in $N$.

The most naive factoring method would be just divide
$N$ by each number from
$1$ to $\sqrt{N}.$ This requires at least $\sqrt{N}$ operations.
Since
$\sqrt{N}=2^{\frac{1}{2}\log N}$ is exponential in the number of digits
 $n=\log N$ in $N$ this method is not an
efficient algorithm. There is no known efficient classical algorithm for
factoring but the quantum polynomial algorithm does exist.

The best classical factoring algorithm which is currently
known is the number field
sieve \cite{Len}. It requires asymptotically
$$
\exp(cn^{1/3}(\log n)^{2/3})
$$
operations for some constant $c$, i.e. it is exponential in $n^{1/3}$.
 P. Shor \cite{Sho1} has found a quantum
algorithm which takes asymptotically
$$
O(n^2\log n\log \log n)
$$
i.e. only a polynomial number
of operations on a quantum computer along with a polynomial amount of
 time on a classical computer.

In these lectures an exposition of Shor`s quantum algorithm for factoring
integers is given together with a short introduction to quantum
computing and number theory. In the description of Shor`s algorithm we
essentially follow his original presentation \cite{Sho1},
see also \cite{EJ}.

It is known that  using randomization the factorization of 
$N$ can be reduced
to finding the {\it order} of an arbitrary element $m$ in
the multiplicative group of
residues modulo $N$; that is the least integer $r$ such that
$$
m^r\equiv 1~(\text{mod}~ {N})
$$
The reduction will be discussed below in Sect.9.
Therefore to factorize $N$ it is enough to find the order $r$ of $m$.

Shor`s algorithm for finding the order consists of 5 steps:

1. Preparation of quantum  state.

2. Modular exponentiation.

3. Quantum Fourier transform.

4. Measurement.

5. Computation of the order at the classical computer.

These steps will be discussed in details.
In Sections 2 and 3 elementary notions of theory of algorithms and
quantum computing are discussed. In particular
a general notion of algorithm
is formulated. In Sect.4 the quantum Fourier transform
is considered.
In Sect.5
some relevant results of number theory are collected.
In Sect. 6 the modular exponentiation is considered.
In Sect.7 Shor`s algorithm for
finding the order is exposed. In Sect. 8 the computational complexity of
Shor`s algorithm is considered. Finally in Sect. 9  the reduction of
problem of factorization to finding the order is discussed.

The main results of the quantum algorithm for finding the order
are given in Theorem 7.1  on the lower  bound for the probability
of
 measurement and in Theorem 7.2 on the derivation of the order. Theorem 8.1
 describes  the computational
complexity of the
 algorithm.  The main result of the quantum algorithm for factoring is
 presented in Theorem 9.2.

\section{Algorithms}
\label{s-alg}

Algorithm is a precise formulation of doing something. Algorithms play an
important role in mathematics and in computers. Algorithms are employed to
accomplish specific tasks using data and instructions. The notion of
algorithm is an old one, there is for example the well known Euclid's
algorithm for finding the greatest common divisor of two numbers. Let us
exhibit Euclid's algorithm here.

\textbf{Euclid's algorithm. }\textrm{Given two positive integers }$m$ and $%
n, $ find their greatest common divisor, i.e. the largest positive integer
which divides both $m$ and $n.$ Here $m$ and $n$ are interpreted as
variables which can take specific values. Suppose that $m$ is greater then $%
n.$ The algorithm consists from three steps.

\textrm{Step 1}\textbf{.}\textrm{\ Divide }$m$ by $n$ and let $r$ be the
remainder$.$

\textrm{Step 2}\textbf{.}\textrm{\ If }$r=0,$ the algorithm halts; $n$ is
the answer.

\textrm{Step 3}\textbf{.}\textrm{\ Replace the value of variable }$m$ by the
current value of variable $n$, also replace the value of variable $n$ by the
current value of variable $r$ and go back to Step 1.

An algorithm has \textit{input}, i.e., quantity which is given to it
initially before the algorithm begins. In Euclid's algorithm the input $%
\mathrm{is}$ a pair of two positive integers $m$ and $n.$ An algorithm has
\textit{output, }\textrm{i.e., quantity which has a specified relation to
the input. }In Euclid's algorithm the output is $n$ in Step 2, which is the
greatest common divisor of two given integers.

{\bf Exercise.}
Prove that the output of Euclid's algorithm is indeed the greatest common
divisor.

Hint: After Step 1, we have $m=kn+r,$ for some integer $k.$
Euclid`s algorithm is considered below in Sect. 3.

There are various approaches to precise formulation of the concept
of algorithm. There exist classical and quantum algorithms. One of
modern precise formulations of the notion of classical algorithm
can be given by using Turing machines. Another approach to
algorithms is based on the notion of circuits. Classical circuits
and classical Turing machines are used as mathematical models of
classical computer. {\it Quantum circuits and quantum Turing machines}
are mathematical models of quantum computer. These important notions were
introduced by D. Deutsch \cite{Deu1,Deu2}.

{\bf Turing Machine.}

The concept of the Turing machine was introduced by A.M. Turing in
1936 for
the study of limits of human ability to solve mathematical problems in
formal way. Any reasonable classical algorithm can be
implemented on a Turing machine
(this is the so called A.Church thesis).

A Turing machine has two main parts:
a \textit{tape} and a central unit with
a\textit{\ head} $\nabla $ (see Figure below).

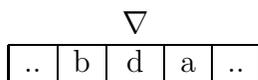
\[
\begin{tabular}{ccccc}
&  & $\nabla $ &  &  \\ \hline
\multicolumn{1}{|c}{..} & \multicolumn{1}{|c}{b} & \multicolumn{1}{|c}{d} &
\multicolumn{1}{|c}{a} & \multicolumn{1}{|c|}{..} \\ \hline
\end{tabular}
\]

The tape is infinite in both directions and is divided into squares. Each
square of the tape holds exactly one of the symbols from a finite set of
symbols (a finite set of symbols is called an alphabet). The central unit
with the head is in one of states from a finite set of states. The head sees
at any moment of time one square of the tape and is able to read the\textrm{%
\ \ }content of the square as well as to write on the square. The input is
written as a string (sequence) of symbols on the tape. The head starts in a
prescribed state. In a single move, the Turing machine can read the symbol
on the one square seen by its head, and based on that symbol and its current
state, replace the symbol by a difference one, change its state, and move
the head one square to the left, or one square to the right, or stays on the
same square as before.

A sequence of moves is called a computation. For some pairs of states and
symbols on the tape the machine halts. In this case, symbols remaining on
the tape form the output, corresponding to the original input. A Turing
machine accepts some input strings if it halts on it. The set of all
accepted strings is called a language accepted by the Turing machine. Such
languages are called recursively enumerable sets.

The Turing machine is a suitable model for the computational power of a
classical computer. Its usefulness follows from {\it the Church's thesis}
 which
may be reformulated as follows: The computational power of the Turing
machine represents a limit for any realizable classical computer.
The Turing machine is considered for example in \cite{GJ}.
 
{\bf General Notion of Algorithm.}

Let us indicate now one method which is general enough to include classical
as well as quantum algorithms. Let us take two sets $I$ and $O.$ The set $I$
will represent input and the set $O$ represents output of our computation.
Suppose the sets $I$ and $O$ are parts of a larger set $\mathsf{S}$ which
will represent configurations of computation. Let $G=\left\{
g_{1},...,g_{r}\right\} $ be a finite set of functions $g_{i}$ from $\mathsf{%
S}$ to $\mathsf{S}.$ Such functions are called \textit{gates }$\mathrm{in}$
computing and $G$ is called the basis of gates. They form the primitive
elements from which we will design an algorithm. For example the gates can
represent the basic logical operations $AND,$ $OR$ and $NOT.$ Now let us be
given a function $f$ which maps the input set $I$ to output set $O.$ Our
problem is to find a sequence of gates $A=\left\{
g_{i_{1}},g_{i_{2}},...,g_{i_{k}}\right\} $ which computes the function $f$
in the sense that the function can be represented as a composition of gates,
i.e. for any input $x\in I$ one has $f(x)=g_{i_{1}}g_{i_{2}}...g_{i_{k}}(x).$
The sequence $A$ is called the {\it algorithm}
or the program of computation.

Each input $x$ in the set $I$ defines a computational sequence, $%
x_{0},x_{1},...$ as follows: $x_{0}=x,$ $%
x_{1}=g_{i_{1}}(x_{0}),...,x_{m}=g_{i_{m}}(x_{m-1}),....$ One says that the
computational sequence \textrm{terminates} in $k$ steps if $k$ is the
smallest integer for which $x_{k}$ is in $O,$ and in this case it produces
the output $y=x_{k}$ from $x.$ One says that the algorithm computes the
function $y=f(x).$

A more general approach would be if one admits that the functions $g_{i}$
and the function $f$ are not defined everywhere (such functions are called
partial functions) and that not every computational sequence terminates.
Moreover one can assume that the transition $x_{m}=g_{i_{m}}(x_{m-1})$ takes
place with a certain probability (random walk) and that the output space $O$
is a metric space with a metric $\rho .$ Then one says that the algorithm
makes an approximate computation of a function $f(x)$ with a certain
probability if one gets a bound $\rho (f(x),x_{k})<\varepsilon .$

To summarize, the algorithm for the computation of the function $f$ by using
the prescribed set of gates is given by the  data $\left\{ \mathsf{S%
},I,O,G,A,f\right\} $ described above.

The set $\mathsf{S}$ for the classical
Turing machine will be the set of all configurations of the Turing machine
and the gates $g_{i}$ form the transition function. For a classical circuit
the gates might be for example basic logical operations $AND,$ $OR$ and $%
NOT. $ For quantum circuit and for quantum Turing machine the set $\mathsf{S}
$ might be the Hilbert space of quantum states and the gates $g_{i}$ could
be some unitary matrices and projection operators.

An important issue in computing is the computational complexity. One would
like to minimize the amount of time and memory needed to produce the output
from a given input. For input $x$ let $t(x)=k$ be the number of steps until
the computational sequence terminates. The computational time $T$ of the
algorithm is defined by
\[
T(n)=\max_{x}\left\{ t(x):\mid x\mid =n\right\}
\]
where $\mid x\mid $ is the length of the description of $x.$
The actual
length of the description depends on the model of computation.

 For input $x$ let $s(x)$ be the number of different elements in the
computational sequence $x_0=x, x_1,...$. The computational space $S $
of the algorithm is defined by
$$
S(n)=\max_{x}\left\{ s(x): s(x) =n\right\}
$$
We are
interested, of course, to minimize the computational time $T(n)$
and space $S(n)$.

\section{Quantum Circuits}\label{s-qc}

{\bf Quantum Mechanics.}

Quantum mechanics was created by W. Heisenberg and E. Schrodinger in 1925.
Together with relativity theory it is the most fundamental theory
in physics. There are two
important points in quantum mechanics:
\begin{itemize}
\item
Quantum mechanics is a statistical theory.
\item
Every quantum system assigns a Hilbert space.
\end{itemize}

Vectors in the Hilbert space represent states of the quantum system,
while self-adjoint operators represent observables. We will need 
only a finite
dimensional Hilbert space which is the $n$-dimensional vector space
$\mathbb{C}^n$ with the scalar product
$$
(z,w)=\sum_{i=1}^n \bar{z}_iw_i
$$
If $\psi$ and $\phi$ are two vectors of the unit length then the
probability to observe the state $\psi$ given the state $\phi$ is
$|(\psi,\phi )|^2$.

{\bf Boolean Functions.} Quantum circuits are quantum analogues of
the classical circuits computing Boolean functions. The Boolean
function $f(x_1,...,x_n)$ is a function of $n$ variables where
each variable takes values $x_i=0,1$ and the function also takes
values 0 and 1. If we denote $B=\{0,1\}$ then the function $f$ is
a map $f:B^n\to B$. One considers also more general Boolean
functions $f: B^n\to B^m$. A classical circuit can be represented
as a directed acyclic graph. Similarly a quantum circuit is a
sequence of unitary matrices of the special form associated with a
(hyper)graph. We will need a special computational basis in the
vector space.

\textbf{Computational basis in }$n-$ \textbf{qubit space}.

 The
two-dimensional complex space $\mathbb{C}^{2}$ is called \textit{qubit. }We
define in qubit the following \textit{computational basis}

\[
e_{0}=\left(
\begin{array}{l}
1 \\
0
\end{array}
\right) ,\qquad e_{1}=\left(
\begin{array}{l}
0 \\
1
\end{array}
\right)
\]

The index $x=0,1$ in the basis $\left( e_{x}\right) $ will be interpreted as
a Boolean variable. We will use also the Dirac notations

\[
e_{x}=\mid x>.
\]

The $n-$tuple tensor product of qubits
$\mathbb{C}^{2}\otimes $ $\mathbb{C}%
^{2}...\otimes $ $\mathbb{C}^{2}=$ $\mathbb{C}^{2^{n}}$ is called the $n-$ qubit
space. It has a \textit{computational basis } $\left\{ e_{x_{1}}\otimes
e_{x_{2}}\otimes ...\otimes e_{x_{n}}\right\} $ where $x_{i}=0,1.$ We will
use also the notation

\[
e_{x_{1}}\otimes e_{x_{2}}\otimes ...\otimes e_{x_{n}}=\mid
x_{1},...,x_{n}>.
\]

If $\psi$ is a vector of the unit length in $\mathbb{C}^{2^{n}}$
then the probability to observe the Boolean variables $x_1,...,x_n$
in the state $\psi$ is
$$
|(e_{x_{1}}\otimes e_{x_{2}}\otimes ...\otimes e_{x_{n}},\psi)|^2
$$
By using the Dirac notations one can write this expression also as
$$
|<x_n,...,x_1\mid \psi>|^2.
$$

{\bf Definition.}
\textit{A quantum circuit }$Q$ is defined by the following set of data:
$$
Q=\left\{ \mathcal{H},U,G\right\}
$$
where the Hilbert space $\mathcal{H}$ is the $n-$ qubit space
$\mathcal{H}=\mathbb{C}^{2^{n}}$, $U$ is a unitary matrix in 
$\mathcal{H},$ and
$G=\left\{ V_{1},...,V_{r}\right\} $ is a finite set of unitary matrices
(quantum gates). The matrix $U$ should admit a 
representation as a product of unitary
matrices generated by the quantum gates described below (\ref{qc2}).

The dimension of unitary matrices $V_{i}$ normally is less then the
dimension $2^{n}$ of the Hilbert space $\mathcal{H}$ and usually one takes
matrices $V_{i}$ which act in the $2-$ qubit or in the $3-$ qubit spaces. We
fix the computational basis $\left\{ e_{x_{1}}\otimes e_{x_{2}}\otimes
...\otimes e_{x_{n}}\right\} $ in $\mathcal{H}$ and define an extension of
the matrix $V_{i}$ to a matrix in the space $\mathcal{H}.$ The extension is
constructed in the following way. If $V_{i}$ is an $l\times l$ matrix then
we choose $l$ vectors from the computational basis and denote them as $%
\alpha =\left\{ h_{1},...,h_{l}\right\} $. Now let us define a unitary
transformation $V_{i}^{(\alpha )}$ in the Hilbert space $\mathcal{H}$ as
follows. The action of $V_{i}^{(\alpha )}$ on the subspace of $\mathcal{H}$
spanned by vectors $\left\{ h_{1},...,h_{l}\right\} $ we set to be equal to $%
V_{i}$ and the action of $V_{i}^{(\alpha )}$ on the orthogonal subspace to
be equal to $0.$

The matrix $U$ should be represented in the following product form

\begin{equation}
U=V_{i_{1}}^{(\alpha _{1})}V_{i_{2}}^{(\alpha _{2})}...V_{i_{L}}^{(\alpha
_{L})}  \label{qc2}
\end{equation}

where the matrices $V_{s}$ are quantum gates and $V_{s}^{(\alpha
_{s})}$ is some extensions of $V_{s}$ to a matrix in the Hilbert
space $\mathcal{H}$ described above.

{\bf Quantum Gates.}

Consider the following set of unitary matrices
$$
G=\{V_1,V_2\}
$$
where $V_1$ is the $2\times 2 $ matrix of rotations to an irrational angle
$\theta$ and $V_2$ is the $4\times 4$ matrix acting to the basis in
$\mathbb{C}^2\otimes\mathbb{C}^2$ as
$$
V_2|x,y>=|x,x+y~(\text{mod}~2)>
$$ where $x,y=0,1.$ The matrix $V_2$ is called the CNOT-operation.
The matrices $V_1$ and $V_2$ gives an example  of universal quantum gates.
By using these gates one can construct a unitary matrix of the form
(\ref{qc2}) which is close as we wish to any unitary matrix in
$\mathbb{C}^{2^{n}}$.

{\bf Exercise.} Let $S_{\theta}=\{e^{2\pi i \theta n}\}$ 
be a set of points
on the unit circle. Here $\theta$ is a fixed irrational number and
$n=0,\pm 1,\pm 2,...$ Prove that the set $S_{\theta}$ is 
a dense set on the unit circle. 
  
Let $f$ be a classical Boolean function $f:B^{k}\rightarrow
B^{m}$. Here $B=\left\{ 0,1\right\} $ and one assumes $k\leq n$ and $m\leq
n.$ 
We say that the \textit{quantum circuit }$Q$\textit{\ computes the Boolean
function} $f:B^{k}\rightarrow B^{m}$ if the following bound is valid

\[
\mid <\mathbf{0},f(x_{1},...,x_{k})\mid U\mid x_{1},...,x_{k},\mathbf{0}%
>\mid ^{2}\geq 1-\varepsilon
\]

for all $x_{1},...,x_{k_{1}}$ and some fixed $0\leq \varepsilon <1/2.$ Here $%
\mid x_{1},...,x_{k},\mathbf{0}>$ is the vector for the computational basis
of the form $\mid x_{1},...,x_{k},0,...,0>$ ($n-k$ zeros) and $<\mathbf{0}%
,f(x_{1},...,x_{k})\mid $ is the vector for the computational basis of the
form $<0,...,0,f(x_{1},...,x_{k})\mid $ ($m-k$ zeros).

If there is a quantum circuit $Q$ with the unitary operator $U$ represented
as a product of $L$ unitary matrices (gates) in the form
(\ref{qc2}) then $L$ is
called
 the \textit{computational time }
of the quantum circuit. We are mainly
interested in the studying of the dependence of $L$ on the 
length of input $k.$

There are different quantum circuits for size of input. Hence actually
we deal with families of quantum circuits.
The computational power of a family of quantum circuits
should be equivalent to quantum Turing machine. This is provided by the
requirement of {\it uniformity}. A family of quantum
circuits is called uniform if its design is produced by a polynomial time
classical computer and if the entries in the unitary matrices of the
quantum circuits are computable numbers.

 For more details about models of quantum computations see for example
\cite{BV,Vol1}.

\section{Quantum Fourier Transform}\label{s-fourier}

Consider the Hilbert space
$\mathbb{C}^{2}\otimes $ $\mathbb{C}
^{2}...\otimes $ $\mathbb{C}^{2}=$ $\mathbb{C}^{2^{s}}$
of the dimension $q=2^s$. Quantum Fourier transform is the unitary
transformation $F_q$ which acts to the computational basis as
\[
F_q|a>=\frac{1}{\sqrt q}\sum_{b=0}^{q-1}e^{2\pi iab/q}|b>
\label{Fou}
\]
Here
$$
|a>=|a_{s-1},...,a_0>,\quad |b>=|b_{s-1},...,b_0>
$$
where one has the binary representations
$$
a=a_0+a_1 2+...+a_{s-1}2^{s-1},~~a_i=0,1
$$
$$
b=b_0+b_1 2+...+b_{s-1}2^{s-1},~~b_i=0,1
$$

{\bf Example. Hadamard`s Gate.}

For $L=1$ the quantum Fourier transform is called the
{\it Hadamard gate},
$F_2=H$. It acts to the basis as
$$
H|0>=\frac{1}{\sqrt 2}(|0>+|1>),
$$
$$
H|1>=\frac{1}{\sqrt 2}(|0>-|1>).
$$
We extend the action  of the Hadamard gate to the $s$-qubit space
as
$$
H_j=I\otimes ...\otimes H\otimes ...\otimes I,~~j=1,2,...,s.
$$

The quantum Fourier transform is multiplication by an $q\times q$
unitary matrix, where the $x,y$ matrix element is $e^{2\pi ixy/q}$.
Naively, this multiplication requires $O(q^2)$ elementary operations.
However, we will show that due to special properties of the quantum Fourier transform,
it can be implemented asymptotically
by means only $O((\log q)^2)$ elementary operations.

It is important to notice that the action of the
quantum Fourier transform can be written in the factorized
(unentangled) form:
$$
F_{2^s}|a_{s-1},...,a_0>=\frac{1}{\sqrt {2^s}}(|0>+e^{i\phi_a 2^{s-1}}|1>)
\otimes (|0>+e^{i\phi_a 2^{s-2}}|1>)\otimes ...\otimes
(|0>+e^{i\phi_a} |1>)
$$
where $\phi_a =2\pi a/2^s$.

We will prove that the quantum Fourier transform can be written
as a product of matrices generated by Hadamard`s gates and 
by the following
$4\times 4$ matrix $B$,
$$
B|a_1,a_0>=\begin{cases}
             e^{i\pi/2}|a_1,a_0>, & \text{if}~ a_1=a_0=1,\\
             |a_1,a_0>, & \text{otherwise}.
            \end{cases}
$$
We denote $B_{j,k},~j<k$ the following extension of the matrix $B$:
$$
B_{j,k}|a_{s-1},...,a_k,...,a_j,...,a_0>=e^{i\theta_{k-j}}
|a_{s-1},...,a_k,...,a_j,...,a_0>
$$
where
$$
e^{i\theta_{k-j}}=\begin{cases}
               (e^{i\pi/2})^{(k-j)}, & \text{if}~ a_1=a_0=1,\\
             1, & \text{otherwise}.
            \end{cases}
$$
The computational complexity of the quantum Fourier transform
is described by the following theorem.

{\bf Theorem 4.1.} Quantum Fourier transform in the space
$\mathbb{C}^{2^{s}}$ can be represented as a product of $O(s^2)$
operators $H_j$ and $B_{j,k}$.

Therefore there is a quantum algorithm for implementation
of quantum Fourier transform which is polynomial as the function of the
input size.

{\bf Proof.}
To explain the proof of the theorem we define the {\it reversal} Fourier
transform
\[
F_q^{Rev}|a_{s-1},...,a_0>=\frac{1}{\sqrt q}
\sum_{b=0}^{q-1}e^{2\pi iab/q}|b_0,b_1,...,b_{s-1}>
\label{FouRe}
\]
In particular one has
$$
F_4^{Rev}=H_0B_{01}H_1.
$$
One can prove an important formula
$$
F_{2^s}^{Rev}=H_0B_{0,1}...B_{0,s-1}H_1...B_{s-4,s-3}B_{s-4,s-2}
B_{s-4,s-1}H_{s-3}B_{s-3,s-2}B_{s-3,s-1}H_{s-2}B_{s-2,s-1}H_{s-1}.
$$
In this formula one has $s$ matrices $H_j$ and $s(s-1)/2$
matrices $B_{j,k}$. Now since $F_q=F_q^{Rev}T$ where $T$ is the
transposition operator, the theorem follows.$\Box$

\section{Elements of Number Theory}\label{Numb}

In this section we  collect
 some relevant material from number theory \cite{Vin}.

\textbf{Euclid`s Algorithm. }Given two integers $a$ and $b,$ not both
zero, the \textit{greatest common divisor }of $a$ and $b,$ denoted \textit{%
g.c.d.}$(a,b\mathit{)}$\textit{\ }is the biggest integer\textit{\ }$d$
dividing both $a$ and $b.$  For example, \textit{g.c.d.(}$9,12$\textit{)}$=3.$

There is the well known \textit{Euclid`s algorithm} of finding the greatest
common divisor. It proceeds as follows.

Find \textit{g.c.d.}$(a,b\mathit{)}$ where $a>b>0.$

1) Divide $b$ into $a$ and write down the quotient $q_{1}$ and the remainder
$r_{1}:$
\[
a=q_{1}b+r_{1}, \quad 0 < r_1 <b,
\]
2) Next, perform a second division with $b$ playing the role of $a$ and $%
r_{1} $ playing the role of $b$:
\[
b=q_{2}r_{1}+r_{2}, \quad 0 < r_2 < r_1,
\]
3) Next:
\[
r_{1}=q_{3}r_{2}+r_{3}, \quad 0 < r_3 < r_2.
\]
Continue in this way. When we finally obtain a remainder that divides the
previous remainder, we are done: that final nonzero remainder is the \textit{%
g.c.d. }of $a$ and $b:$
\begin{eqnarray*}
r_{t} &=&q_{t+2}r_{t+1}+r_{t+2}, \\
r_{t+1} &=&q_{t+3}r_{t+2}.
\end{eqnarray*}
We obtain: $r_{t+2}=d=$\textit{g.c.d.}$(a,b\mathit{).}$

{\bf Example}. Find \textit{g.c.d}$.(128,24):$
\begin{eqnarray*}
128 &=&5\cdot 24+8, \\
24 &=&3\cdot 8
\end{eqnarray*}
We obtain that \textit{g.c.d}$.(128,24)=8.$

Let us prove that Euclid`s algorithm indeed gives the greatest
common divisor. Note first that $b>r_{1}>r_{2}>...$ is a sequence
of decreasing positive integers which can not be continued
indefinitely. Consequently Euclid`s algorithm must end.

Let us go up through out Euclid`s algorithm. $r_{t+2}=d$ divides $r_{t+1},$ $%
r_{t},...,r_{1},b,a.$ Thus $d$ is a common divisor of $a$ and $b.$

Now let $c$ be any common divisor of $a$ and $b.$ Go downward
through out Euclid`s algorithm. $c$ divides
$r_{1},r_{2},...,r_{t+2}=d.$ Thus $d,$ being a common divisor of
$a$ and $b,$ is divisible by any common divisor of these numbers.
Consequently $d$ is the greatest common divisor of $a$ and
$b.\Box$

Another (but similar) proof is based on the formula
\[
\mathit{g.c.d}.(qb+r,b)=\mathit{g.c.d}.(b,r).
\]
{\bf Corollary.} Note that from Euclid`s algorithm it follows
 (go up) that if $d=$\textit{%
g.c.d.}$(a,b\mathit{)}$ then there are integers $u$ and $v$ such
that
\begin{equation}
d=ua+vb.  \label{num2}
\end{equation}
In particular one has
\begin{equation}
ua\equiv d~(\text {mod}~ b)  \label{num2a}
\end{equation}

One can estimate the efficiency of Euclid`s algorithm. By \textit{Lame`s
theorem} the number of divisions required to find the greatest common
divisor of two integers is never greater that five-times the number of
digits in the smaller integer.

\textbf{Congruences.} An integer $a$ is \textit{congruent to }$b$ \textit{%
modulo }$m,$
\[
a\equiv b ~(\text{mod}~m)
\]
iff $m$ divides $(a-b).$ In this case $a=b+km$ where $k=0,\pm
1,\pm 2,...$.

{\bf Proposition}. Let us be given two integers $a$
and $m$. The following are equivalent

(i) There exists $u$ such that $au\equiv 1~(\text{mod}~m).$

(ii) $\mathit{g.c.d}.(a,m)=1.$

\textbf{Proof.} From (i) it follows
\[
ab-mk=1.
\]
Therefore the $\mathit{g.c.d}.(a,m)=1,$ i.e. we get (ii).

Now if (ii) is valid then one has the relation (\ref{num2a}) for
$d=1,b=m$:
\[
au\equiv 1~(\text {mod}~m)
\]
which gives (i).$\Box$

Let us solve in integers the equation
\begin{equation}
ax\equiv c~(\text{mod}~ m)  \label{num3}
\end{equation}
We suppose that $\mathit{g.c.d}.(a,m)=1.$ Then by the previous proposition
there exists such $b$ that
\[
ab\equiv 1~(\text{mod}~m).
\]
Multiplying Eq (\ref{num3}) to $b$ we obtain the solution
\begin{equation}
x\equiv bc~(\text{mod}~m)  \label{num4}
\end{equation}
or more explicitly
\[
x=bc+km,\qquad k=0,\pm 1,\pm 2,...
\]
{\bf Exercise.} Find all of the solutions of the congruence
$$
3x\equiv 4~(\text{mod}~7).
$$
{\bf Continued Fractions.}

Euclid`s algorithm is closely related with continued fractions.
If $a$ and $b$ are two integers then by using Euclid`s algorithm we write
$$
a=q_{1}b+r_{1}; \quad \frac{a}{b}=q_1+\frac{1}{b/r_1},
$$
$$
b=q_{2}r_{1}+r_{2}; \quad \frac{b}{r_1}=q_2+\frac{1}{r_1/r_2},
$$
$$
r_{1}=q_{3}r_{2}+r_{3}; \quad \frac{r_1}{r_2}=q_3+\frac{1}{r_1/r_2},
$$
$$
.......................................
$$

\begin{eqnarray*}
r_{t} &=&q_{t+2}r_{t+1}+r_{t+2}; \quad  \frac{r_t}{r_{t+1}}=
q_{t+2}+\frac{1}{r_{t+1}/r_{t+2}},\\
r_{t+1} &=&q_{t+3}r_{t+2}; \quad \frac{r_{t+1}}{r_{t+2}}=q_{t+3}.
\end{eqnarray*}
Therefore we obtain a representation of $a/b$ as a continued fraction
$$
\frac{a}{b}=q_1+\cfrac{1}{q_2+
                 \cfrac{1}{q_3+...
                  \cfrac{1}{q_{t+3}
                  }}}
$$
Hence any positive rational number can be
represented by a continued fraction.
Fractions
$$
\delta_1=q_1,\quad \delta_2=q_1+\frac{1}{q_1},\quad \delta_3=
q_1+\frac{1}{q_2+\frac{1}{q_3}},...
$$
are called convergents. We will use the following

{\bf Theorem 5.1}. If $x$ is a rational number and $a$
 and $b$ are positive integers satisfying
$$
|\frac{a}{b}-x|<\frac{1}{2b^2}
$$
then $a/b$ is a convergent of the continued fraction of $x$.

\textbf{Chinese Remainder Theorem.} Suppose there is a system of
congruences to different moduli:
\begin{eqnarray*}
x &\equiv &a_{1}~(\text{mod}~{m_1}), \\
x &\equiv &a_{2}~(\text{mod}~m_{2}), \\
&&... \\
x &\equiv &a_{t}~(\text{mod}~ m_{t})
\end{eqnarray*}
Suppose  $\mathit{g.c.d}.(m_{i},m_{j})=1$ for $i\neq j.$ Then there exists a
solution $x$ to all of the congruences, and any two solutions are congruent
to one another modulo
\[
M=m_{1}m_{2}...m_{t}.
\]
\textbf{Proof. }Let us denote $M_{i}=M/m_{i}.$ There exist $N_{i}$ such that
\[
M_{i}N_{i}\equiv 1~(\text{mod}~m_{i})
\]
Let us set
\[
x={\sum }_i a_{i}M_{i}N_{i}
\]
This is the solution. Indeed we have
\[
{\sum }_i a_{i}M_{i}N_{i}=a_{1}M_{1}M_{1}+...\equiv
a_{1}+a_{2}+...\equiv a_{1}~(\text{mod}~m_{1})
\]
and similarly for other congruences.$\Box$

We will need also

{\bf Fermat`s Little Theorem}.
Let $p$ be a prime number. Any integer $a$ satisfies
\[
a^{p}\equiv a~(\text{mod}~p)
\]
and any integer $a$ not divisible by $p$ satisfies
\[
a^{p-1}\equiv 1~(\text{mod}~p).
\]
\textbf{Proof.} Suppose $a$ is not divisible by $p$. Then
$\{0a,1a,2a,...,(p-1)a\}$ form a complete set of residues modulo $p$,
i.e. $\{a,2a,...,(p-1)a\} $ are a rearrangement of $\{1,2,...,p-1\}$
when considered modulo $p$. Hence the product of the numbers in the first
sequence is congruent modulo $p$ to the product of the members
in the second sequence, i.e.
$$
a^{p-1}(p-1)\equiv (p-1)!~(\text{mod}~p)
$$
Thus $p$ divides $(p-1)(a^{p-1}-1)$. Since $(p-1)!$ is not
divisible by $p$, it should be that $p$ divides
$(a^{p-1}-1)$.$\Box$

{\bf The Euler function.}

The Euler function $\varphi(n)$ is the number of nonnegative integers $a$
less then $n$ which are prime to $n$:
$$
\varphi(n)=\#\{0\leq a < n:g.c.d.(a,n)=1\}
$$
In particular $\varphi (1)=1,~\varphi (2)=1,...,\varphi
(6)=2,...$. One has $\varphi (p)=p-1$ for any prime $p$.

{\bf Exercise.} Prove: $\varphi (p^n)=p^n-p^{n-1}$ for any $n$ and prime $p$.

The Euler function is multiplicative, meaning that
$$
\varphi (mn)=\varphi (m)\varphi (n)
$$
whenever $g.c.d.(m,n)=1$.

If
$$
n=p_1^{\alpha_1}p_2^{\alpha_2}...p_k^{\alpha_k}
$$
then
$$
\varphi (n)=n(1-\frac{1}{p_1})...(1-\frac{1}{p_k})
$$
In particular, if $n$ is the product of two primes, $n=pq$, then
$$
\varphi (n)=\varphi (p)\varphi (q)=(p-1)(q-1)
$$
There is the following generalization of Fermat`s Little Theorem.

{\bf Euler`s theorem.} If $g.c.d.(a,m)=1$ then
$$
a^{\varphi (m)}\equiv 1~(\text{mod}~m).
$$
{\bf Proof.} Let $r_1,r_2,...,r_{\varphi (m)}$
be classes of integers relatively
prime to $m$. Such a system is called a reduced system of residues
mod $m$. Then $ar_1,ar_2,...,ar_{\varphi (m)}$ is another reduced system
since $g.c.d.(a,m)=1$. Therefore
$$ar_1\equiv r_{\pi (1)},~ar_2\equiv r_{\pi (2)},...,ar_{\varphi (m)}\equiv
r_{\pi (m)}~(\text{mod}~m)
$$
On multiplying these congruences, we get
$$
a^{\varphi (m)}r_1r_2...r_{\varphi (m)}\equiv
r_1r_2...r_{\varphi (m)}~(\text{mod}~m)
$$
Now since $r_1r_2...r_{\varphi (m)}$ is relatively prime to $m$
the theorem is proved.$\Box$

We will use the following result on the asymptotic behaviour of the
Euler function.

{\bf Theorem 5.2.} There is a constant $C>0$ such that for sufficiently large
$n$ one has
\[
\frac{\varphi (n)}{n}~\geq ~\frac{C}{\log\log n}.
\]

\section{Modular Exponentiation}\label{s-modexp}

Sometimes it is necessary to do classical computations on quantum
computer. Since quantum computation is reversible, a deterministic
classical computation is performable on quantum computer only if
it is reversible. It was shown that any deterministic computation
can be made reversible for only a constant factor cost in time and
by using as much space as time.

In this section we discuss the modular exponential problem.
The problem is, given $N,a$ and $m, ~m\leq N,~ a\leq N$ find
$m^a~(\text{mod}~N)$.

{\bf Theorem 6.1.}
There exists a classical algorithm for computation
$m^a~(\text{mod}~N)$ which requires
asymptotically $O(n^2 \log n \log \log n)$ arithmetical
operations with bits in the binary representation of the numbers
where $n=\log N$.

{\bf Proof.} The algorithm proceeds as follows.

1. Write the binary representation
$$
a=a_0+2a_1+2^2a_2+...+2^sa_s
$$ where $a_i=0,1$ and $a_0=1$.

2. Set $m_0=m$ and then for $i=1,...,s$ compute
$$
m_i\equiv m_{i-1}^2 m^{a_{s-i}}~(\text{mod}~N)
$$

3. The final result is $m_s$ ,
$$
m_s=m^a~(\text{mod}~N)
$$

The validity of the algorithm follows from the relation
$$
m_i\equiv m^{a_0+2a_1+...2^ia_i}~(\text{mod}~N)
$$

Computation at the third step requires no more then three
multiplication and it is repeated no more then $s\leq n=\log N$
times. There is the Sch$\ddot{o}$nhage-Strassen algorithm
\cite{SS} for integer multiplication that uses asymptotically
$O(n\log n\log\log n)$ operations on bits. This proves the
theorem. $\Box$

Note that the Sch$\ddot{o}$nhage-Strassen algorithm
is the best known algorithm for
multiplication of the very large numbers but for intermediate length
numbers (several thousand digits)
it might be better to use the original Karatsuba algorithm
\cite{Kar} which requires $O(n^{\log_2 3})$ operations,
$n=\log N$.

\section{Shor`s Algorithm for Finding the Order}\label{SAlgOr}

Given $N$ choose a random (with the uniform distribution ) $m$
$(1<m\leq N)$. We assume $gcd (m,N)=1$, otherwise we would already know a
divisor of $N$.
We want to find the order of $m$, i.e. the least integer $r$ such that
$$
m^r\equiv 1 ~(\text{mod}~ N)
$$
Fix some $q$ of the form $q=2^s$ with $N^2\leq q<2N^2$.
The algorithm will use the Hilbert space
$$
{\cal H}=\mathbb{C}^{q}\otimes \mathbb{C}^{N_1}
\otimes \mathbb{C}^{k}
$$
where $\mathbb{C}^{q}$ and $\mathbb{C}^{N_1}$
are two quantum registers which hold integers represented in binary.
Here $N_1$ is an integer of the form $N_1=2^l$ for some $l$ such that
$N\leq N_{1}$. There is also the work space $\mathbb{C}^{k}$
to make arithmetical
operations. We will not indicate it explicitly.
If
$$
a=a_0+2a_1+2^2a_2+...+2^sa_s
$$
is the binary representation ($a_i=0,1$) of an integer $a$ then
we write
$$
|a>=|a_0>\otimes ... \otimes |a_s>
$$
where
$$
|0>=\left(
\begin{array}{l}
1 \\ 0
\end{array}
\right) ,\qquad |1>=\left(
\begin{array}{l}
0 \\ 1
\end{array}\right)
$$
is the basis in the two dimensional complex space $\mathbb{C}^2$.

We have the data $(N,m,q)$. The algorithm for finding the order $r$
of $m$ consists from 5 steps:

1. Preparation of quantum  state.

2. Modular exponentiation.

3. Quantum Fourier transform.

4. Measurement.

5. Computation of the order at the classical computer.

{\bf Description of the algorithm}.

{\bf 1. Preparation of quantum  state}. Put the first register in the uniform
superposition of states representing numbers $a~~(\text{mod}~q)$.
The quantum
computer will be in the state
$$
|\psi_1>=\frac{1}{\sqrt q}\sum_{a=0}^{q-1}|a>\otimes |0>
$$

{\bf 2. Modular exponentiation}. Compute $m^a~(\text{mod}~N)$
in the second register.
This leaves the quantum computer in the state
$$
|\psi_2>=\frac{1}{\sqrt q}\sum_{a=0}^{q-1}|a>\otimes |m^a~(\text{mod}~N)>
$$

 {\bf 3. Quantum Fourier transform}. Perform the quantum Fourier transform on the
first register, mapping $|a>$ to
$$
\frac{1}{\sqrt q}\sum_{c=0}^{q-1}e^{2\pi iac/q}|c>
$$
The quantum computer will be in the state
$$
|\psi_3>=\frac{1}{q}\sum_{a=0}^{q-1}\sum_{c=0}^{q-1}e^{2\pi iac/q}|c>
\otimes |m^a~(\text{mod}~N)>
$$

{\bf 4. Measurement}. Make the measurement on both registers $|c>$ and
$|m^a \pmod{N}>$.

To find the period $r$ we will need only the value of
$|c>$ in the first register but for clarity of computations we make the
measurement on the both registers.
The probability $P(c,m^k~(\text{mod}~ N))$
that the quantum computer ends in a particular state
$$
|c;m^k~( \text{mod}~N)>=|c>
\otimes |m^k~( \text{mod}~(N)>
$$
is
\begin{equation}
 P(c,m^k~(\text{mod}~ N))=|<m^k ~(\text{mod}~N);c|\psi_3>|^2
\label{P()}
\end{equation}
where we can assume $0\leq k<r$.

We will use the following Theorem which shows that the probability
$P(c,m^k~(\text{mod}~N))$ is large if the residue of 
$rc~(\text{mod}~q)$ is small.
Here $r$ is the order of $m$ in the group  $(Z/NZ)^*$
of residues of modulo $N$.

{\bf Theorem 7.1}. If there is an integer $d$ such that
\begin{equation}
-\frac{r}{2}\leq rc-dq\leq\frac{r}{2}
\label{rc}
\end{equation}
and $N$ is sufficiently large then
\begin{equation}
\label{Pla}
P(c,m^k~(\text{mod}~N))\geq\frac{1}{3r^2}
\end{equation}
The theorem is proved below.

{ \bf 5. Computation of the order at the classical computer}. We
know $N,c$ and $q$
and we want to find the order $r$. Because $q>N^2$, there is at most one
fraction $d/r$ with $r<N$ that satisfies the inequality (\ref{rc}). 
We can
obtain the fraction $d/r$ in lowest terms by rounding $c/q$ to the nearest
fraction having a denominator smaller than $N$. To this end we can use
the continued fraction expansion of $c/q$ and Theorem 5.1.

 We will prove the following
theorem which summarizes  main results of the quantum algorithm for
finding the order.

{\bf Theorem 7.2.} If the integer $N$ is sufficiently large then
by repeating the first four steps of the algorithm for finding the
order $O(\log\log N)$ times one can obtain the value of the order
$r$ with the probability $\gamma >0$ where the constant $\gamma$
does not depend on $N$.

Now let us prove these results.

{\bf Proof of Theorem 7.1.} First let us notice the relation
$$
<m^k ~(\text{mod}~N)|m^a ~(\text{mod}~N)>=
\begin{cases}
             1, & \text{if}~ a\equiv k~(\text{mod}~r),\\
             0, & \text{otherwise}.
            \end{cases}
$$
Hence the amplitude
$$
<m^k ~(\text{mod}~N);c|\psi_3>=
\frac{1}{q}\sum_{a}e^{2\pi iac/q}
$$
where the summation on $a$ runs on the subset $ a\equiv k~(\text{mod}~r)$
of the set $\{0,1,...,q-1\}$.
One sets
$$
a=br+k$$
to get
$$
\sum_{a}e^{2\pi iac/q}=\sum_{b=0}^f e^{2\pi ic(br+k)/q}=
 \frac{1-e^{2\pi icr(f+1)/q}}{1-e^{2\pi icr/q}}e^{2\pi ick/q}
$$
where $f$ is the integer part
$$
f=\left [\frac{q-1-k}{r}\right ]
$$
Therefore the probability is
$$
P(c,m^k~(\text{mod}~N))=
\left |\frac{1}{q}\frac{1-e^{2\pi icr(f+1)/q}}
{1-e^{2\pi icr/q}}\right |^2
$$
which is equal to
$$
P(c,m^k~(\text{mod}~N))=\frac{1}{q^2}\frac{\sin^2 \frac{\pi cr(f+1)}{q}}
{\sin^2\frac{\pi cr}{q}}
$$
Now to prove the theorem we use the condition (\ref{rc}) and the relation
$$
\sin x > \frac{2}{\pi}x, \quad 0 < x < \frac{\pi}{2}. \quad \Box
$$

{\bf Proof of Theorem 7.2.} If we know the fraction $d/r$ in lowest terms
and if $d$ is relatively prime to $r$ then we can derive $r$.
There are $r\varphi (r)$ states $|c;m^k~( \text{mod}~N)>$ which enable us
to compute $r$ because there are $\varphi (r)$ values of $d$
relatively prime to $r$ and also there are $r$ possible values
for $m^k~(\text{mod}~N)$. By Theorem 7.1 
each of these states occurs with the probability
at least $1/3r^2$. Therefore we can get $r$ with probability
at least $\varphi (r)/3r$. Now the theorem follows from 
Theorem 5.2.$\Box$


\section{Computational Complexity of Shor`s Algorithm}\label{ComCompl}

Let us estimate the number of operations (or gates) needed to implement the
first three steps of the Shor`s algorithm for finding the order.

{\bf Theorem 8.1.} Shor`s algorithm for finding the order of an element in the
group  of residues of modulo $N$ requires
\begin{equation}
O((\log N)^2(\log\log N)(\log\log\log N))
\label{ord9}
\end{equation}
operations (gates) at a quantum computer.

{\bf Proof.} Let us estimate the number of operations (gates)
needed to implement the
first three steps of the algorithm at a quantum computer.

To prepare the state $|\psi_1>$ one needs
$$
s=\log q=O(\log N)
$$
Hadamard`s gates.

Then let us consider the modular exponentiation. It is the most
time consuming part of the algorithm. As it is discussed in Sect.
6, asymptotically, modular exponentiation requires
\begin{equation}
O(n^2\log n\log \log n)
\label{mod9}
\end{equation}
operations, $n=O(\log N)$. The computation can be made reversible
for only a constant factor cost in time and the same amount in
space.

Finally, it is shown in Sect. 4 that to make the third step of the
algorithm, quantum Fourier transform, one takes
\begin{equation}
O((\log N)^2)
\label{key}
\end{equation}
quantum gates. Actually this is the key ingredient in the
factoring algorithm. Just because of the polynomial bound
(\ref{key}) we  obtain the polynomial efficiency of the factoring
algorithm.

Now the theorem follows from estimates (\ref{ord9}),(\ref{mod9})
and (\ref{key}).$\Box$


\section{Factoring Integers}\label{fact}

In this Section the factoring algorithm will be described.
The factoring algorithm solves the following problem.
 Given an integer $N,$ one
has to find   such integers $p$
and $q$ that $N=pq$ or to prove that such a factoring does not exist.
It is assumed that $p$ and $q$ are not equal to 1.
We shall use the algorithm for finding the order described in Sect.7.

{\bf Factoring Algorithm.}

1. Choose a random $m,~1\leq m\leq N$ (with uniform distribution)
and find its order $r$
by using the factoring algorithm from Sect.7.

2. If $r$ is even, compute
$$
g.c.d.(m^{r/2}-1,N)
$$
by using Euclid`s algorithm.

3. If $g.c.d.(m^{r/2}-1,N)>1$ then it gives a factor of $N$.
In the case if $g.c.d.(m^{r/2}-1,N)=1$ or the order $r$ of $m$
is odd one has to repeat the steps 1 and 2 for another integer $m$.

Let us explain why the algorithm works. Consider equation
$$
y^2\equiv 1~(\text{mod}~N)
$$
There are trivial solutions
$$
y\equiv \pm 1~(\text{mod}~N)
$$
Suppose there is also a nontrivial solution $y=b$,
$$
b^2\equiv 1~(\text{mod}~N); \quad b\not\equiv \pm 1~(\text{mod}~N)
$$
Then
$$
(b+1)(b-1)\equiv 0~(\text{mod}~N)
$$
i.e.
$$
(b+1)(b-1)=kN
$$
and neither of the factors $b+1$ and $b-1$ is $0~(\text{mod}~N)$.
Thus, $(b+1)$ must contain one factor of $N$ and $(b-1)$ another.

Now, if $r$ is the order of $m ~(\text{mod}~N)$ and $r$ is even, then
$b=m^{r/2}$ is the solution of equation
$
b^2\equiv 1~(\text{mod}~N)
$.
If $m^{r/2}\not\equiv\pm 1~(\text{mod}~N)$ then
$g.c.d.(m^{r/2}-1,N)>1$. We have proved the following

{\bf Lemma.}  If the order $r$ of $m~(\text{mod}~N)$ is even and
$$
m^{r/2}\not\equiv\pm 1~(\text{mod}~N)
$$
then
$$
g.c.d.(m^{r/2}-1,N)>1
$$

 The above process may fail if $r$ is odd or if $r$ is even but
$m^{r/2}$ is a trivial solution. However, due to the following theorem,
these situations can arise only with small probability.

{\bf Theorem 9.1.} Let $N$ be an odd natural number with prime
factorization
$$
N=p_1^{\alpha_1}p_2^{\alpha_2}...p_k^{\alpha_k}
$$
Suppose $m$ is chosen at random, $1\leq m\leq N$ (with uniform
distribution),
satisfying  $g.c.d.(m,N)=1$. Let $r$ be the order of $m~(\text{mod}~N)$.
Then
\begin{equation}
Prob~\{r:r \text{~is even and}~ m^{r/2}\not\equiv\pm 1~(\text{mod}~N)\}\geq
1-\frac{1}{2^{k-1}}
\end{equation}
The probability is positive if $k\geq 2$.

{\bf Proof.} Since $r$ is the order we never have 
$m^{r/2}\equiv -1~(\text{mod}~N)$. One can prove that
$$
Prob~\{r:r \text{~is odd or}~ m^{r/2}\equiv -1~(\text{mod}~N)\}
\leq \frac{1}{2^{k-1}}
$$
by using the Chinese remainder theorem.$\Box$

 {\bf Theorem 9.2.}  If an integer $N$ is sufficiently
large and if it is a product of at least two prime numbers then
the factoring algorithm finds the factors with the probability
greater then $\gamma/2$ where $\gamma$ is the constant defined in
Theorem 7.2. One needs asymptotically
$$
O((\log N)^2 (\log\log N)(\log\log\log N))
$$
quantum gates to implement the quantum circuit for the factoring
algorithm.

 {\bf Proof.} The conclusion of the theorem follows
from the description of the factoring algorithm and from Theorems
7.2 and 9.1.$\Box$

\section{Conclusions}
Factoring integers plays an important role in modern cryptography, see
\cite{VV}. This explains the significance of Shor`s quantum factoring
algorithm.

There are important problems such as the traveling salesman
problem, the integer programming problem, the satisfiability
problem that have been studied for decades and for which all known
algorithms have a running time that is exponential in the length of
the input. These problems and many other problems belong to the
set of $NP$-complete problems. It is unknown whether the factoring
problem is $NP$-complete. Probably it is only subexponential in
the running time.

An approach to the solution of $NP$-complete problems 
by using a new paradigm of computations which
goes beyond the quantum Turing machine is suggested in
\cite{OV2}. It is based on
a combination of quantum computer with a chaotic dynamics amplifier
and on an application of nonlinear
Hartree-Fock dynamics in atomic quantum computer \cite{Vol2}.

\section*{Acknowledgments}

This work was supported in part by RFFI 99-0100866
 and by INTAS 99-00545 grants.


\end{document}